\documentclass[a4paper,onecolumn,11pt,unpublished]{quantumarticle}
\pdfoutput=1
\usepackage[utf8]{inputenc}
\usepackage[english]{babel}
\usepackage[T1]{fontenc}
\usepackage{amsmath}
\usepackage{hyperref}

\usepackage{amssymb}

\usepackage{tikz}
\usepackage{lipsum}

\begin{document}

\title{Floquet engineering of quantum thermal machines: A gradient-based spectral method to optimize their performance}

\author{Alberto Castro}
\affiliation{Departamento de F{\'{\i}}sica Te\'orica, At\'omica y \'Optica, University of Valladolid, Spain}
\affiliation{Institute for Biocomputation and Physics of Complex Systems (BIFI) of the University of Zaragoza, Spain}
\orcid{0000-0002-9253-7926}
\email{alberto.castro.barrigon@gmail.com}
\maketitle

\begin{abstract}
A procedure to find optimal regimes for quantum thermal engines (QTMs)
is described and demonstrated. The QTMs are modelled as the periodically-driven non-equilibrium
steady states of open quantum systems, whose dynamics is approximated in this work
with Markovian master equations. The 
action of the external agent and the couplings to the heat reservoirs can be modulated with
control functions, and it is the time-dependent shape of these control functions the object
of optimisation. Those functions can be freely parameterised, which permits to
constrain the solutions according to experimental or physical requirements.
\end{abstract}


\section{Introduction}
\label{sec:intro}

\begin{figure}
\centerline{\includegraphics{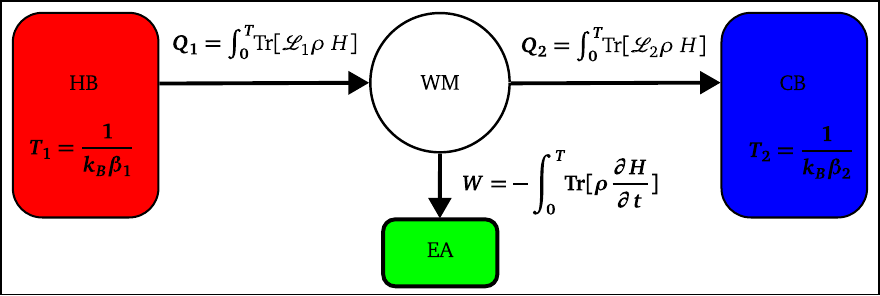}}
\caption{\label{fig:diagram} Basic diagram of a QTM: a hot bath (HB)
  at temperature $T_1$ and a cold batch (CB) at temperature $T_2$
  exchange heat with the working medium (WM). This can also exchange
  work with an external agent (EA). The equations are the expressions
  for the heats and work assuming a Markovian master equation (see
  text).  The directions of the arrows suggest a heat engine operation
  mode: $Q_1 > 0$ (the WM receives heat from the hot bath), $Q_2 < 0$
  (the WM gives away part of that heat to the cold bath), and $W > 0$
  (the remaining heat is transformed into output work on the EA).  }
\end{figure}

Thermal machines are devices composed of a {\it working fluid} (or
{\it working medium}), one or more {\it heat reservoirs}, and an {\it
  external agent}. The heat reservoirs or {\it baths} are macroscopic
systems, typically at thermal equilibrium, and normally large enough
so that one can assume that they are not altered by their interaction
with the working fluid. The working fluid itself may be any substance
capable of exchanging energy with the reservoirs in the form of
heat. Furthermore, the working fluid exchanges energy with the
external agent in the form of work -- either performed on or by the
working fluid.  Depending on the sign and relative values of those
heats and works, the thermal machine is a {\it heat engine}, a {\it
  refrigerator}, a {\it heat pump}, etc. The basic setup
  is sketched in Fig.~\ref{fig:diagram}.

Historically, the theory of thermodynamics was developed around the
analysis of the thermal machine. It was well established way before
quantum mechanics, but the laws of equilibrium thermodynamics
cannot be considered ``classical'' or ``quantum'', as the
theory is by definition agnostic about the microscopic dynamics of the
constituents of the system that it studies. It is however normally
assumed that the systems are macroscopic in size: it is a theory about
systems ``in the thermodynamic limit''.

But this need not be the case for thermal machines, and Scovil and
Schulz-Dubois showed in 1959~\cite{Scovil1959} how a three-level maser can be
analysed as a quantum thermal machine (QTM)~\cite{Bhattacharjee2021,
  Alicki2015}. The road toward the miniaturisation of all kinds of devices that has been followed
in the last decades has raised the interest in analysing micro and
mesoscopic systems as tentative working mediums. Numerous proposals
for QTMs have been put forward, often only as theoretical proposals,
but also as experimental realisations~\cite{Blickle2012, Abah2012,
  Rossnagel2016, vonLindenfels2019, Peterson2019, Klatzow2019,
  Passos2019}.

A fairly large body of literature on the topic of QTMs and, in
general, of quantum thermodynamics~\cite{Binder2018, Cleri2024,
  Kosloff2013} has been produced in the last decades. Unsurprisingly,
the topic of optimal efficiencies and bounds or limits for output
powers and performances has often been investigated, given that the
bound for the efficiency of a heat engine established by Sadi Carnot
is perhaps the most popular formulation of the II law of
thermodynamics~\cite{Carnot1824}. In fact, the seminal paper of Scovil and
Schulz-Dubois~\cite{Scovil1959} found that the maser efficiency is
also bound by the value predicted by Carnot.

The theoretical absolute limits for the performance of these machines
are however unattainable in practice and,
moreover, they may require useless operation modes. For example, the
paradigmatic limit of classical thermodynamics, Carnot's efficiency,
can only be reached assuming that the thermal machine is evolving {\em
  quasistatically}, i.e. staying at all time
in equilibrium, which essentially means infinitely slow. Therefore,
the output power per unit time of a heat engine performing Carnot's
cycle is zero. This regime is both unattainable and useless from a
practical perspective, hence the need for working with {\em finite
  time} thermodynamics. In this realm, the dynamics of the microscopic
constituents of the systems cannot be ignored any more -- such as it
is in the pure field of equilibrium thermodynamics --, and one can
start to wonder about differences between the quantum and classical
cases.

Erdman {\em et al.}~\cite{Erdman2019} also found
that the Carnot efficiency limit can be reached for QTM machines
modelled as two-level-systems (TLS) with tunable gap: they demonstrated
how the optimal regime is in this case found with infinitely fast
two-stroke Otto cycles (switching very rapidly from a large gap when
the system is coupled to the the hot bath, to a smaller gap when the
system is coupled to the cold bath). It is clear how the Carnot limit
is only achieved (or, one should say, approached to arbitrary
precision) with experimentally impossible requirements:
infinitesimally short strokes, and sudden, discontinuous Hamiltonian
changes.

The research on the theoretical absolute bounds for the performances
of QTMs has been extensive over the last decades.  However, it is also
important to develop techniques for the more mundane computational
task of finding optimal protocols when using experimentally realistic
external agents and control handles. A number of works have addressed
this more practical issue: for example, reinforcement learning has
very recently been used for this task~\cite{Erdman2022, Erdman2023,
  Erdman2023b}. Machine learning (in this case, deep learning) was
also proposed by Khait {\em et al.}~\cite{Khait2022}.
  Normally, these methods based on machine learning are {\em
    gradient-free}: they do not employ the gradient of the merit
  function that is to be optimised with respect to the control
  variables.

However, this type of optimization problem also belongs to the
class of problems addressed by (quantum) optimal control theory
(Q)OCT~\cite{Kirk1998, Brif2010, Glaser2015, Koch2022}: finding the
control functions that maximise a merit function of the evolution of
the state. Surprisingly, only a few works have used this method:
perhaps the most remarkable is the work of Cavina and
collaborators~\cite{Cavina2018, Cavina2018b}, who made an explicit use
of Pontryagin's maximum principle (PMP)~\cite{Pontryagin1962,
  Boscain2021}, the standard workhorse of OCT.

In contrast to the methods based on machine learning mentioned above,
QOCT based on the PMP approaches the problem examining the gradient --
although, most often, the gradient is properly speaking a functional
derivative, as the control variables are normally control {\em
  functions}. The functional derivative of the merit function with
respect to the control functions has to be zero at an optimal control -- a
condition that can be formulated as a set of nonlinear equations, as
stated by the PMP. Those equations can sometimes be solved directly, or the
gradient of functional derivative can be used to feed an optimization
algorithm leading to its nullification.

Notice that, in
purity, the optimization of QTMs working in cycles, i.e. {\em
  periodically}, should be addressed by {\em periodic} optimal
control, a subclass of OCT that has received less attention.
One technique for dealing with periodic systems and working
on their optimisation with also periodic control functions is the
pseudospectral Fourier approach (see for example
Ref.~\cite{Elnagar2005}).

In this work, I propose to explore that path: to develop
  a method to perform optimisations on QTMs by recasting the master
  equations that describe their evolution, assumed to be Markovian, in
  the Fourier domain. It builds on the method already described in
Ref.~\cite{Castro2023} to optimise averaged values of observables for
driven periodic non-equilibrium steady states of open quantum
systems. However, it needs to be generalised to account for more
general observables (transferred heats and averaged output powers). In
Ref.~\cite{Castro2023} we used the term ``Floquet
engineering''~\cite{Oka2019}, which has been coined in the last
decades to refer to the manipulation of materials through the use of
periodic perturbations. Recently, this author and collaborators have
shown one possible method to couple this concept with OCT (see, for
example~\cite{Castro2022, Castro2023-2}; other methods have been
proposed, see for example ~\cite{Verdeny2014, Petiziol2021,
  Kalinowski2023} in the field of quantum simulators). The work
described below extends this concept to QTMs modelled as open quantum
systems, and therefore it can be termed as Floquet engineering of
QTMs. The method essentially consists in parameterising the control
functions according to the experimental or physical requirements, and
working out a computationally feasible expression for the gradient of
the target or merit function with respect to those parameters. This
gradient may then be used to feed any maximisation algorithm.

Section~\ref{sec:qtms} summarises some key concepts about QTMs in order
to set the frame and notation used in this
article. Section~\ref{sec:floquetengineering} describes the technique
used to optimise their performance. Section~\ref{sec:examples}
describes some examples of optimisations and, finally,
Section~\ref{sec:conclusion} presents the conclusions of the
work. Hereafter, we will assume $\hbar = 1$ and $k_{\rm B} = 1$.


\section{Quantum thermal machines as periodically driven non-equilibrium steady states}
\label{sec:qtms}

The suitable framework to describe the operation of QTMs is the theory
of open quantum systems~\cite{Breuer2007, Kosloff1984}. In this
framework, the working fluid is the only piece of a QTM that is
explicitly accounted for; the heat reservoirs constitute the
environment that is factored out, whereas the external agent that
gives or receives work is only included as a normally time-dependent
part of the Hamiltonian of the working fluid. Hereafter, we will
furthermore assume the Markovian approximation, which can be used if
certain conditions are met: essentially, the reservoir correlation
times must be much shorter than the relaxation time of the system, and
the system-bath interactions must be weak. The most general form for
the equation of motion of an open system -- the so-called master
equation -- in the Markovian approximation was demonstrated to be the
Gorini-Kossakowski-Sudarshan-Lindblad (GKSL) form~\cite{Lindblad1976,
  Gorini1976}. In its original formulation, it accounted only for
static Hamiltonians, but it can be extended to the time-dependent
case. 
Here, we are concerned with time-periodic GKSL equations:
\begin{equation}
\dot{\rho}(t) = \mathcal{L}(\underline{f}(t))\rho(t)\,.
\end{equation}
The Lindbladian time-dependence is assumed to be determined by the
functions $\underline{f}(t)$, a set of $m$ time-periodic functions of time,
\begin{equation}
f_k(t+T)=f_k(t),\;(k=0,1,\dots,m-1)\,,
\end{equation}
with period $T$, that permit to control the precise form of the
Lindbladian $\mathcal{L}$.  The system is in contact with a number
(normally, two) of heat reservoirs at different temperatures, and
therefore we split $\mathcal{L}$ as:
\begin{equation}
\mathcal{L}(\underline{f}(t)) = \mathcal{L}_{H}(\underline{f}(t)) + \sum_b \mathcal{L}_b(\underline{f}(t))\,,
\end{equation}
where
\begin{equation}
\mathcal{L}_H(\underline{f}(t))X = -i \left[ H(\underline{f}(t)),X\right]
\end{equation}
is the unitary or coherent part of the time-evolution generator,
whereas each $\mathcal{L}_b(\underline{f}(t))$ is an incoherent operator that determines the interaction of the
system with reservoir $b$.

The Hamiltonian $H(t)$ may include a Lamb shift; otherwise, it is simply
the Hamiltonian that would generate the isolated evolution (see \cite{Correa2024} for a
recent discussion on the necessity of including Lamb shift and renormalization terms
when deriving Markovian master equations). Some of the terms of $H(\underline{f}(t))$
may be time-dependent, controlled by some of the functions $\underline{f}(t)$:
those are the drivings originated by the external agent. Likewise, the
interaction between the system and the reservoirs may also depend on
some of the functions $\underline{f}(t)$, allowing for example for the switching
on and off of cold or hot baths, etc.

In the presence of both the periodic drivings and of the baths, under
rather general assumptions~\cite{Menczel2019}, the system will
eventually decay into a periodic NESS:
\begin{equation}
\rho_{\rm NESS}(t+T) = \rho_{\rm NESS}(t)
\end{equation}
(in the following, the ``NESS'' label will not be explicitly used, as all density
matrix trajectories $\rho(t)$ will correspond to a NESS).
This can then be viewed as a quantum thermal machine that
performs a cycle of period $T$, giving and receiving energy into and
from the baths ({\em heat}), and giving or receiving energy into and from the
source of the external driving ({\em work}).

The energy balance can be understood in terms of those
concepts. Defining the instantaneous energy function as:
\begin{equation}
E(t) = {\rm Tr}\left[\rho(t)H(\underline{f}(t))\right]\,,
\end{equation}
we must have, in the NESS, $E(T) = E(0)$.
Following Alicki~\cite{Alicki1979}, the variation of this energy can be broken down as:
\begin{equation}
\frac{{\rm d}E}{{\rm d}t}(t) = -p(t) + \sum_b j_b(t)\,,
\end{equation}
where:
\begin{align}
j_b(t) &= {\rm Tr}\left[\mathcal{L}_b(\underline{f}(t))\rho(t)\;H(\underline{f}(t))\right]\,,
\\
p(t) &= -{\rm Tr}\left[\rho(t)\frac{\partial H}{\partial t}(\underline{f}(t))\right]\,.
\end{align}
These are the energy flows transferred to the system, per unit time,
from the baths and to the external agent, i.e. the transferred heats
and work, respectively (or, if the sign is negative, energies per unit
time transferred {\em to} the baths or {\em from} the external agent). One may
then define the amounts of heats and work over one cycle:
\begin{align}
Q_b = \int_0^T\!\!{\rm d}t\; j_b(t)\,,
\\
W = \int_0^T\!\!{\rm d}t\; p(t)\,.
\end{align}
These terms are sketched in Fig.~\ref{fig:diagram}.
We will use these energies per unit time (i.e. with dimensions
of {\em power}), $J_b = Q_b/T$, and $P = W/T$. Given the periodic
behaviour of our system,
\begin{equation}
\int_0^T\!\!{\rm d}t\; \frac{{\rm d}E}{{\rm d}t}(t) = E(T)-E(0) = 0\,,
\end{equation}
we must have an energy balance that is usually presented as the
formulation of the I Law of thermodynamics for QTMs:
\begin{equation}
P = \sum_b J_b\,.
\end{equation}

\section{Floquet-engineering QTMs}
\label{sec:floquetengineering}

The goal now is to find those control functions $f$ that lead the QTM
to work in an optimal regime. The definition of what ``optimal'' means
may of course vary. For example, one may wish to maximise the power
output of a quantum engine, its efficiency, or the coefficient of
performance of a refrigerator. In general, the goals would probably be
functions of the energy terms $J_b$ and $P$ defined above.

Rather than working with unconstrained functions of time, it is more
convenient to parameterise these functions,
\begin{equation}
f_k = f_k(\underline{u}^{(k)}, t)\quad (k=0,\dots,m-1)\,,
\end{equation}
where each $\underline{u}^{(k)}$ is a set of {\em control parameters}, that we
collectively group into $\underline{u}$ to ease the notation, as we collectively
group all $f_{k}$ into the multi-dimensional function $\underline{f}$. In this
way, it is much easier to constrain the functions to experimentally or
physically meaningful forms (in terms of frequencies, amplitudes,
etc.)  Therefore, the task is to find the optimal set of control
parameters $\underline{u}^{\rm opt}$ that lead to forms for the functions $\underline{f}$ that
optimise the machine behaviour.

We will hereafter denote $\rho(\underline{u}, t)$ to the periodic solution (NESS)
of the master equation:
\begin{align}
\label{eq:leq1}
\dot{\rho}(\underline{u}, t) &= \mathcal{L}(\underline{f}(\underline{u}, t))\rho(\underline{u}, t)\,.
\\
\label{eq:leq2pc}
\rho(\underline{u}, t+T) &= \rho(\underline{u}, t)
\end{align}

The optimisation problem must be formulated by first establishing the goal: a
functional of the behaviour of the system during one cycle,
\begin{equation}
F = F(\rho, \underline{u})\,,
\end{equation}
where the $\rho$ dependence refers to the full periodic trajectories in the cycle. 
The extra dependence on $\underline{u}$ may be used to add penalties over undesirable regions
of parameter space (an example of this will be given later).

The goal is therefore to maximise function
\begin{equation}
G(\underline{u}) = F(\rho(\underline{u}, \cdot), \underline{u})\,,
\end{equation}
where now $\rho(\underline{u}, \cdot)$ denotes the particular periodic trajectory that is the NESS
solution to Eqs.~(\ref{eq:leq1}) and (\ref{eq:leq2pc}).

In order to solve this optimisation problem, the first ingredient is
therefore a computational procedure to obtain the NESS $\rho(\underline{u},
\cdot)$, and function $G(\underline{u})$ from it. Numerous optimisation methods
exist that permit to obtain optimal values for functions with only
that ingredient. However, more effective methods can be used if one
also has a procedure to compute the gradient of $G$. By applying
the chain rule for functional derivatives in order to get an expression
for this gradient,
\begin{align}
\nonumber
\frac{\partial G}{\partial u_r}(\underline{u}) &= \frac{\delta F}{\delta \rho}(\rho(\underline{u}, \cdot), \underline{u})
\left( \frac{\partial \rho}{\partial u_r}(\underline{u}, \cdot)\right)
\\
\nonumber
& + \frac{\delta F}{\delta \rho^*}(\rho(\underline{u}, \cdot), \underline{u})
\left( \frac{\partial \rho^*}{\partial u_r}(\underline{u}, \cdot)\right)
\\
\label{eq:gradient-G}
& + \frac{\partial F}{\partial u_r}(\rho(\underline{u}, \cdot), \underline{u})\,,
\end{align}
it becomes clear that the second necessary ingredient for the optimisation of $G$ 
involves the computation of the gradient of $\rho(\underline{u}, \cdot)$ with respect to
the control parameters $\underline{u}$.

In Ref.~\cite{Castro2023}, we demonstrated the feasibility of a
computational procedure to obtain these derivatives, and consequently,
the feasibility of a procedure for the optimisation of function
$G$. In that work, it was limited to functionals $F$ defined as
averages of observables, i.e.:
\begin{equation}
F(\rho, \underline{u}) = \frac{1}{T}\int_0^T\!\!{\rm d}t\; {\rm Tr}\left[A\rho(\underline{u}, t)\right]\,,
\end{equation}
although it can be extended to more general cases -- for
example, functions of the heats and power flowing to and from a QTM,
as it will shown below.

Let us start by briefly summarising the procedure used in \cite{Castro2023} to obtain
the NESS and its gradient with respect to the control parameters. The starting point are the periodic Lindblad
equations (\ref{eq:leq1}-\ref{eq:leq2pc}) in the frequency domain -- a transformation using Fourier series that
will automatically imply the periodicity of all the objects:
\begin{equation}
\sum_\beta\sum_{p=0}^{N-1}\left[ \mathcal{L}_{\alpha\beta,q-p}(\underline{u})
-i\delta_{pq}\delta_{\alpha\beta}\omega_p\right]\tilde{\rho}_{\beta,p}(\underline{u}) = 0\quad (q = 0, 1,\dots,N-1).
\end{equation}
Here, $\omega_p = \frac{2\pi}{T}p\;(p \in \mathbb{Z})$ are the Fourier expansion frequencies,
$N$ is the integer that sets a cutoff for the Fourier expansion, and
\begin{eqnarray}
\tilde{\rho}_{\beta, p}(\underline{u}) 
&=& \frac{1}{T}\int_0^T\!\!{\rm d}t\; e^{-i\omega_p t}\rho_{\alpha}(\underline{u}, t),
\\
\mathcal{L}_{\alpha\beta, q}(\underline{u}) 
&=& \frac{1}{T}\int_0^T\!\!{\rm d}t\; e^{-i\omega_q t}\mathcal{L}_{\alpha\beta}(\underline{u}, t),
\end{eqnarray}
are the Fourier coefficients of the elements of the density matrix and
Lindbladian. Note that we are using here a vectorized representation of the
density (a vector in Liouville space): the indices $\alpha$ or $\beta$ run over the $d^2$ elements of the density matrix ($d$ being the dimension of the underlying Hilbert space of the working fluid). The Lindbladian is then a rank two operator in Liouville space or {\em superoperator} and requires two indices, $\alpha\beta$.

By further defining
\begin{equation}
\overline{\mathcal{L}}_{\alpha q,\beta p}(\underline{u}) = 
\mathcal{L}_{\alpha\beta,q-p}(\underline{u}) -i\delta_{qp}\delta_{\alpha\beta}\omega_p\,,
\end{equation}
we finally arrive to:
\begin{equation}
\label{eq:homogeneous}
\sum_\beta \sum_{p=0}^{N-1} 
\overline{\mathcal{L}}_{\alpha q,\beta p}(\underline{u}) \tilde{\rho}_{\beta, p}(\underline{u}) = 0\,,
\end{equation}
or
\begin{equation}
\label{eq:homogeneous2}
\overline{\mathcal{L}}(\underline{u})\tilde{\rho}(\underline{u}) = 0\,.
\end{equation}
in matrix form. Note that the dimension of vector $\tilde{\rho}$ is $d^2N$, and the operator
$\overline{\mathcal{L}}(u)$ is a $d^2N\times d^2N$ matrix.

This is a linear homogeneous equation; the solution
(the nullspace or kernel, assuming that it has dimension one), will be
the periodic solution that we are after, the NESS.
We now need some procedure to find $\frac{\partial \rho}{\partial u_r}$.
Taking variations of Eq.~(\ref{eq:homogeneous}) with respect to the parameters $\underline{u}$, we get:
\begin{equation}
\label{eq:mainequation}
\overline{\mathcal{L}}(\underline{u})\frac{\partial \tilde{\rho}}{\partial u_r}(\underline{u}) =
- \frac{\partial \overline{\mathcal{L}}}{\partial u_r}(\underline{u})\tilde{\rho}(\underline{u}).
\end{equation}
This is a linear equation that would provide $\frac{\partial
  \tilde{\rho}}{\partial u_r}$. However, note that since $\overline{\mathcal{L}}(\underline{u})$ has
a non-empty kernel (given precisely by $\tilde{\rho}(\underline{u})$), it cannot be solved
straightforwardly. In fact, it does not have a unique solution: If $x$
is a solution of
\begin{equation}
\overline{\mathcal{L}}(\underline{u})x = 
- \frac{\partial \overline{\mathcal{L}}}{\partial u_r}(\underline{u})\tilde{\rho}(\underline{u}),
\label{eq:compute-x}
\end{equation}
$x + \mu \tilde{\rho}(\underline{u})$ is also a solution for any $\mu$.
To remove this arbitrariness, we impose the normalisation condition,
${\rm Tr}\rho(\underline{u}) = 1$ for any $\underline{u}$, and therefore:
\begin{equation}
{\rm Tr} \frac{\partial \rho}{\partial u_r} = 0.
\label{eq:condition-for-gradient}
\end{equation}
To find $\frac{\partial \rho}{\partial u_r}$ in practice, one may
then take the following two steps: First, compute a solution of the
linear equation, Eq.~(\ref{eq:compute-x}), with the least-squares
method, by imposing that the solution $x_0$ is perpendicular to the
kernel, i.e.: $x_0^\dagger \cdot \tilde{\rho}(\underline{u}) = 0$. Then, update the
solution with the condition,
Eq.~(\ref{eq:condition-for-gradient}). The required solution is
obtained as:
\begin{equation}
\frac{\partial \rho}{\partial u_r} = x_0 - ({\rm Tr} x_0) \rho(\underline{u}).
\end{equation}
Once we have $\frac{\partial \rho_u}{\partial u_r}$, we can evaluate the gradient 
in Eq.~(\ref{eq:gradient-G}). Armed with this procedure to compute this gradient,
one can perform the optimisation of function $G(\underline{u})$ with many efficient algorithms.
This method has been implemented in the qocttools code~\cite{Castro2024}, publicly
available, and all the necessary scripts and data necessary to replicate the following results
are also available upon request from the authors.

As for possible choices for the function $G(\underline{u})$,
for the purposes of this work, we are concerned with target goals
defined in terms of either the averaged power $P$ or the heats $J_b$ (or
combinations of those). For example, if the goal is to maximise the output power of a heat engine,
\begin{align}
\nonumber
G(\underline{u}) = P(\underline{u}) &= -\frac{1}{T}\int_0^T\!\!{\rm d}t\;
{\rm Tr}\left[ \frac{\partial H}{\partial t}(\underline{f}(\underline{u}, t)) \rho(\underline{u}, t)\right]\,,
\\
\label{eq:pdef}
&= -\frac{1}{T}\sum_k \int_0^T\!\!{\rm d}t\; \dot{f}_k(\underline{u}, t) 
{\rm Tr}\left[ V_k(\underline{f}(\underline{u},t)) \rho(\underline{u}, t)\right]\,.
\end{align}
Note the negative sign due to the convention used for the definition of the power $P$.
Here, we use the notation $\dot{f}_k(\underline{u}, t)$ for the time
derivative of function $f_k(\underline{u}, t)$, and
\begin{equation}
V_k = \frac{\partial H}{\partial f_k}\,.
\end{equation}

One must now work out the gradient of this function, for example
making use of the chain rule (\ref{eq:gradient-G}), plugging
the gradient $\frac{\partial \rho}{\partial u_r}$ calculated
with the procedure described above.
But, rather than working out explicitly the
functional derivatives, one may directly work out
the gradient components of function $P(\underline{u})$ from Eq.~(\ref{eq:pdef}):
\begin{align}
\nonumber
\frac{\partial P}{\partial u_r}(\underline{u}) = &
- \frac{1}{T}\sum_k \int_0^T\!\!{\rm d}t\; \left\{ 
\frac{\partial \dot{f}_k}{\partial u_r}(\underline{u}, t)
{\rm Tr}\left[
V_k(\underline{f}(\underline{u},t))\rho(\underline{u},t) 
\right] + \right.
\\
\nonumber
& \sum_l \dot{f}_k(\underline{u}, t) \frac{\partial f_l}{\partial u_r}(\underline{u}, t) 
{\rm Tr} \left[ \frac{\partial V_k}{\partial f_l}(\underline{f}(\underline{u}, t))\rho(\underline{u}, t) \right] +
\\
& \left. 
\dot{f}_k(\underline{u}, t){\rm Tr}\left[ V_k(\underline{f}(\underline{u}, t)) 
\frac{\partial\rho}{\partial u_r}(\underline{u}, t)\right]
\right\}\,.
\end{align}
Despite the length of the equation, in fact the only difficulty lies in computing
the NESS $\rho(\underline{u}, t)$ and its derivatives $\frac{\partial
  \rho}{\partial u_r}$.

A similar procedure can be followed for the case in which function
$G(\underline{u}) = J_b(\underline{u})$, the heat transferred from one of the reservoir. In
the most general case, function $G$ would be a function of all the energy terms,
$G(\underline{u}) = g(P(\underline{u}), J_1(\underline{u}), \dots)$, a function of the power output and of all the
heats (such as the efficiency of a heat engine or the coefficient of
performance of a refrigerator), and then one would have:
\begin{equation}
\frac{\partial G}{\partial u_r} = 
\frac{\partial g}{\partial P} \frac{\partial P}{\partial u_r}
+ \sum_b \frac{\partial g}{\partial J_b} \frac{\partial J_b}{\partial u_r}\,.
\end{equation}
In this way, one can define and optimize a merit function
that combines both the output power and the performance, since normally one would not like to have a very high
power at the cost of a very low performance, or viceversa, but rather have a compromise (see for example \cite{Carrega2024}, a
recent work that studies the Pareto front defined in terms of those two quantities).

However, perhaps one also wishes to include another key quantity, the fluctuations of the output power, as an ingredient 
of the figure of merit -- normally, one would like a heat engine to work with low fluctuations. A combined study
of the optimisation of power, efficiency and fluctuations is for example described in Ref.~\cite{Erdman2023}; see also \cite{Das2023, Campisi2015}. The
present scheme can also be extended to include the fluctuations in the definition of the target; see Appendix
\ref{appendix:fluctuations}.

Finally, a note about the computational complexity of the method: the bottlenecks
are the solution of the linear systems \ref{eq:homogeneous2} and \ref{eq:mainequation}.
The preparation of the matrices and vectors involved in those equations take a comparatively
small amount and time. The dimension of the linear problem is $D = d^2N$, where $d$ is the dimension
of the Hilbert space of the system (a two-level system in this work), and $N$ is the Fourier
decomposition dimension (given by the choice of cutoff). The scaling of a dense linear problem is $D^3$ with
standard methods, and $\approx D^{2.3}$ with more sophisticated methods.



\section{Examples of application}
\label{sec:examples}

\subsection{GKSL equations}

Until now, the form of the master equation has remained rather general
-- although we are always assuming here an important simplification:
the open quantum system is Markovian. Therefore, the equation must be
of the GKSL form~\cite{Lindblad1976, Gorini1976}. The optimisation method
described above may be used for any equation of that family. However,
it has only been implemented and tested for a subclass of GKSL equations:
hereafter, in order to exemplify the method, we will restrict the
analysis to those GKSL equations that verify:

\begin{enumerate}

\item The decoherence terms have the form:
\begin{equation}
\label{eq:decoherenceterm}
\mathcal{L}_b(\underline{f}(\underline{u}, t)) = \sum_i g_{bi}(\underline{f}(\underline{u},t))L_D(\gamma_{bi}, L_{bi})\,.
\end{equation}
where we define the super-operator $L_D(\gamma, X)$ (for any positive constant $\gamma$ and operator $X$) as:
\begin{equation}
L_D(\gamma, X)\rho = \gamma\left(X\rho X^\dagger - \frac{1}{2}\left\{X^\dagger X,\rho\right\}\right)\,.
\end{equation}
Therefore, in this setup, we restrict the Lindblad
operators $L_{bi}$ to be constant in time, but they may be modulated by
time-dependent functions (the so-called ``rates'' may depend on time).

\item We have two reservoirs at thermal equilibrium (as it is almost
always the case): one hot bath $(b=1)$ and one cold bath $(b=2)$.

\item The dependence of $H$ on the control functions is linear, i.e.:
\begin{equation}
H(\underline{f}(\underline{u}, t)) = H_0 + \sum_k f_k(\underline{u}, t)V_k\,.
\end{equation}
and therefore the terms $V_k$ are constant operators, independent of
$\underline{u}$ or time.

\end{enumerate}

This is the type of model that has been implemented in the qocttools
code~\cite{Castro2024} to demonstrate the feasibility of the
optimisation scheme explained above. The key equations are two: on the on hand, the
expression for the gradient, that in this case reduces to:
\begin{equation}
\frac{\partial P}{\partial u_r}(\underline{u}) = 
-\frac{1}{T}\sum_k \int_0^T\!\!{\rm d}t\; \left\{ 
\frac{\partial \dot{f}_k}{\partial u_r}(\underline{u}, t)
{\rm Tr}\left[
V_k\rho(\underline{u},t) 
\right] + 
\dot{f}_k(\underline{u}, t){\rm Tr}\left[ V_k \frac{\partial\rho}{\partial u_r}(\underline{u}, t)\right]
\right\}\,.
\end{equation}
And, in order to find the gradient of $\rho$ 
[Eq.~(\ref{eq:mainequation})], since
\begin{equation}
\frac{\partial \overline{\mathcal{L}}}{\partial u_r} = 
\sum_k \frac{\partial \overline{\mathcal{L}}}{\partial f_k}\frac{\partial f_k}{\partial u_r}\,,
\end{equation}
the key equation is:
\begin{equation}
\frac{\partial \mathcal{L}}{\partial f_k} = -i\left[V_k, \cdot\right]
+ \sum_b\sum_i
 \frac{\partial g_{bi}}{\partial f_k}(\underline{f}(\underline{u}, t)) L_D(\gamma_{bi}, L_{bi})\,.
\end{equation}

\subsection{Model}
\label{subsection:model}

Let us now present the model used for the sample
optimisations shown below. We consider the model used by Erdman {\em
  et al.}~\cite{Erdman2019} to study the optimal Otto cycles (see also
\cite{Erdman2017, Erdman2022, Erdman2023}): a two level system with a
controlled energy gap, i.e.:
\begin{equation}
H(\underline{f}(t)) = \frac{1}{2}(\Delta + f_0(t))\sigma_z\,.
\end{equation}
Note that in this subsection \ref{subsection:model} we are dropping
the dependence on $\underline{u}$ to ease the notation.

Regarding the decoherence terms [Eq.~(\ref{eq:decoherenceterm})],
there are two terms per bath, indexed as $i=+,-$, and
\begin{equation}
L_{b+} = \sigma_+\,,\;\; L_{b-} = \sigma_-\,,
\end{equation}
for both the hot and cold bath ($b=1,2$). All rate constants $\gamma_{bi}$ are
set to be equal ($\gamma_{bi} = \Gamma$), but they are then modulated by
the time-dependent functions
\begin{equation}
g_{bi}(\underline{f}(t)) = f_b(t)F(i\beta_b (\Delta + f_0(t)))\,.
\end{equation}
where $\beta_b$ is the (inverse) temperature of bath $b$, and
\begin{equation}
F(x) = \frac{1}{1+e^x}\,.
\end{equation}
This choice ensures the fulfillment of the detailed balance condition.

Note that we have three {\em control functions}: $f_0(t)$ is
responsible for modifying the TLS gap, whereas $f_1(t)$ and $f_2(t)$
tune the coupling of the system to the hot and cold bath,
respectively.

This model has been used to describe a quantum dot with only one
relevant resonance, coupled to metallic leads with flat densities of
states, that act as reservoirs~\cite{Beenaker1991, Erdman2017,
  Erdman2019, Erdman2022}. Erdman {\em et al.}~\cite{Erdman2019}, in
particular, solved exactly and analytically the following optimisation
problem: suppose that we can vary at will the TLS gap by modulating
$f_0$, as long as a maximum and a minimum are not surpassed: $\vert f_0(t)\vert
\le \delta$. This means there exists a minimum and a maximum TLS gap:
\begin{equation}
\varepsilon_{\rm min}  = \Delta - \delta \le 
\Delta + f_0(t)
\le
\varepsilon_{\rm max}  = \Delta + \delta\,.
\end{equation}
Suppose that we can also vary at will the system-bath coupling functions $f_1$
and $f_2$, as long as $0\le\vert f_b(t)\vert \le 1$. All these control
functions are periodic, with a period $T$ that can also be
varied. Suppose now that we want to optimize the output power of the
QTM operating as heat engine (other possible performance measures were
also considered in \cite{Erdman2019}).

The solution was demonstrated to be the following (see the discussion around the Eq.~(8)
of \cite{Erdman2019}): the maximum is achieved with infinitesimally
short ($T\to 0$) periods, consisting of coupling the system for equal
periods of time ($T/2$) to the hot and the cold baths:
\begin{align}
\label{eq:f1}
f_1(t) = 1\;\; \textrm{and}\;\; f_2(t) = 0\;\;\textrm{if}\;\; 0\le t\le \frac{T}{2}\;\;\;
&\textrm{(coupling to the hot bath)}\,,
\\
\label{eq:f2}
f_1(t) = 0\;\; \textrm{and}\;\; f_2(t) = 1\;\;\textrm{if}\;\; \frac{T}{2}\le t\le T\;\;\;
&\textrm{(coupling to the cold bath)}\,.
\end{align}
During each of those {\em strokes}, the TLS gap has some constant values,
$\varepsilon_1$ and $\varepsilon_2$, respectively. In this setup,
the output power is given by:
\begin{equation}
\label{eq:opsudden}
P_{\rm c}(T, \varepsilon_1, \varepsilon_2) =
\frac{1}{\Gamma \textrm{coth}(\Gamma\frac{T}{4})}
(F(\varepsilon_1 \beta_1) - F(\varepsilon_2 \beta_2))(\varepsilon_1 - \varepsilon_2)\,,
\end{equation}
where the subindex ${\rm c}$ stands for ``constant'', to stress the fact that the function
$f_0(t)$ is constant during each time of contact with the bath: $f_0(t) = \epsilon_b - \Delta$ when
in contact with bath $b$, and it changes value instantaneously when the bath changes.
The value of this output power grows with decreasing periods $T$; in the limit $T\to 0$,
\begin{equation}
\label{eq:absmax}
P_{\rm c}(\varepsilon_1, \varepsilon_2) =
\frac{\Gamma}{4}(F(\varepsilon_1 \beta_1) - F(\varepsilon_2 \beta_2))(\varepsilon_1 - \varepsilon_2)\,.
\end{equation}
The absolute maximum output power for this type of machine is then found at the maximum of this function:
\begin{equation}
\label{eq:opmaxinfintelyshort}
P_{\rm c}^{\rm max} = \max_{\varepsilon_{\rm min} \le \varepsilon_1,\varepsilon_2 \le \varepsilon_{\rm max}} 
P_{\rm c}(\varepsilon_1,\varepsilon_2)\,.
\end{equation}

This can be viewed as a two-strokes engine cycle, that switches
discontinuously from the cold to the hot bath, with no adiabatic
segments. The expansion and compressions (modifications of the TLS
gap, in this case), are instantaneous. Therefore, even at finite $T$,
the operation requires discontinuous jumps in the control functions.

\subsection{Examples of optimizations}

Let us now modify the nature of the problem described above: suppose
that we are not allowed to use a non-smooth control function $f_0$:
the change in time of the TLS gap cannot be sudden, which implies a continuous
and differentiable function $f_0$. We still ask of $f_0$ to be constrained
in amplitude, $\vert f_0(\underline{u}, t)\vert \le \delta$, as mentioned above,
but also demand that it has no frequency components beyond a cutoff
$\omega_{\rm max}$. This cutoff forbids, of course, a sudden discontinuous
change when the system decouples from one bath and couples to the
other one. Furthermore, we fix the cycle period $T$, which in a
realistic setup cannot be taken to arbitrarily close-to-zero values.

The rest of the setup remains unchanged: $f_1$ and $f_2$ are given by
Eqs.~(\ref{eq:f1}) and (\ref{eq:f2}), which means that once again we
have a two-stroke cycle that switches the contact from the hot bath to
the cold bath. We therefore consider these functions to be fixed: they
do not depend on any control parameters $u$ and are not, in purity,
control functions. The optimisation is only done with respect to the
shape of $f_0 = f_0(\underline{u}, t)$ (this is of course not a requirement of the
method, but merely a choice for the examples shown here).

Regarding the parameterisation of $f_0$, it is chosen in such a way
that, by definition, $\vert f_0(\underline{u}, t)\vert \le \delta$ as in the
problem described above. Furthermore, the function is periodic,
continuous and differentiable, and has low frequencies. The detailed
description of the parameterised form of $f_0$ is given in
Appendix~\ref{appendix:parametrization}.

It remains to define the merit function $G$ for this example, which is:
\begin{equation}
G(\underline{u}) = P(\underline{u}) - 
\alpha \sum_{\omega_k > \omega_{\rm max}} \vert \tilde{f}_{0k}(\underline{u})\vert^2\,.
\end{equation}
The goal is therefore to maximise the output power $P(\underline{u})$ as given by
Eq.~\ref{eq:pdef}; but note that we add an extra term: it is a {\em
  penalty} term for high frequencies in the control function
($\tilde{f}_{0k}(\underline{u})$ are the Fourier components of $f_0$). As
discussed in Appendix~\ref{appendix:parametrization}, the
parameterisation forbids amplitudes higher than $\delta$, and {\em
  favours} frequencies lower than $\omega_{\rm max}$, but does not
forbid them. Therefore, in the optimisation we add this extra term to
make them negligible. The constant $\alpha > 0$ determines how
important this penalty is, and therefore how large those frequencies
can be in the resulting optimised function.

Then, the function $P$ and its gradient are computed according to the
formulas described in the previous section, and this information is
fed into an optimisation algorithm. We have chosen the sequential
quadratic programming algorithm for nonlinearly constrained
gradient-based optimisation (SLSQP)~\cite{Kraft1994}, as implemented
in the NLopt library~\cite{NLopt}. This is a versatile choice that
permits to include linear and non-linear bounds and constraints.

For all the calculations shown below, the amplitude constraint is set as
$\delta = (1/5)\Delta$ and the temperatures for the reservoirs are set
to $\beta_1 = 1/\Delta$, $\beta_2 = 2/\Delta$, and the rate $\Gamma = \Delta$, equal for
all the dissipation terms.

\begin{figure}
\centerline{\includegraphics{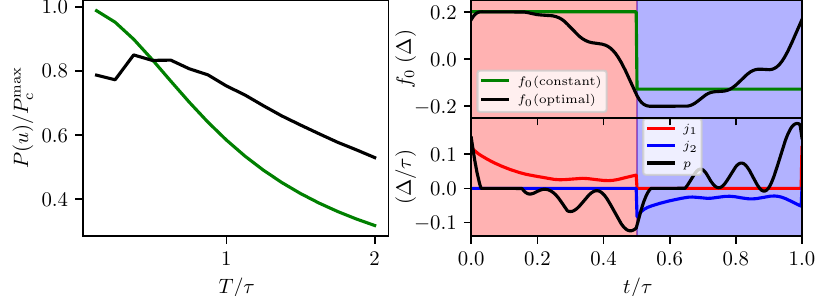}}
\caption{\label{fig:1} {\it Left panel:} Averaged output power of the
  heat engine as a function of the period time $T$, shown as a fraction of the maximum possible power given in Eq.~(\ref{eq:opmaxinfintelyshort}), when using the:
  {\it green}: constant TLS gaps with discontinous fast switching when
  changing bath; and {\it black}: optimized smooth gap $f_0(\underline{u}^{\rm
    opt}, t)$. {\it Top right panel:}
  Function $f_0$ with constant TLS gaps and sudden switching (green),
  and optimized $f_0(\underline{u}^{\rm opt}, t)$. The red and blue shadings mark
  the time regions when the hot and blue baths are connected,
  respectively. {\it Bottom right panel:} Transferred heats and work
  for the heat engine when using $f_0(\underline{u}^{\rm opt}, t)$.  }
\end{figure}

Fig.~\ref{fig:1} displays the first calculation examples. It is a
series of optimisations for varying values of the cycle period $T$,
ranging from $(1/8)\tau$ to $2\tau$, where $\tau =
\frac{2\pi}{\Delta}$. The goal is to optimise the output power
obtained with a protocol $f_0(\underline{u}, t)$ for each of those cycle periods,
and compare that output power with the one that results of using
constant TLS gaps during each contact with the hot and cold bath, with
a sudden, instantaneous change in between. The output power obtained
with those constant gaps is the one obtained by maximising
Eq.~\ref{eq:opsudden} with respect to $\varepsilon_1$ and
$\varepsilon_2$ (within the allowed range $[\Delta -\delta, \Delta +
  \delta]$). The results are shown in the left panel of
Fig.~\ref{fig:1}. The green line displays the output power obtained
with the constant gaps; it can be seen how it increases with
decreasing $T$, and it tends to the maximum predicted by
Eq.~\ref{eq:opmaxinfintelyshort}, as expected.

However, for a fixed and non-zero $T$, the values obtained with
constant gaps are not the largest output powers that one can get; in
order to find the optimal protocol, one must look in the space of
non-constant, varying TLS gaps, for which purpose one has to use a
numerical procedure such as the one proposed in this work. The results
obtained in this way are shown with the black line of the left panel
of Fig.~\ref{fig:1}. It can be seen how, for small $T$, the output
powers are actually lower, and only become larger at a certain crossing
point. The reason is the fact that we are demanding of the protocol to
have frequencies lower than a certain cutoff (which for these examples
we have set to $\omega_{\rm max} = 8\Delta$). It is therefore not
surprising that, for very rapid cycles, the optimised $f_0$ cannot
improve the constant-gap protocol, that approaches the predicted
absolute maximum as $T \to 0$.  For longer cycles, the black curve
does show higher output powers.

The right panel of Fig.~\ref{fig:1} presents the optimal function
$f_0(\underline{u}^{\rm opt}, t)$ (top) and the corresponding transferred heats and
work (bottom) corresponding to the heat engine working with a period
of $T=\tau$. For comparison, the protocol using the optimised constant
gaps is also shown in the top panel (green line). It can be seen how,
as expected, the energy exchange between system and external agent is
higher around the times that the baths are coupled and
decoupled. The optimised function $f_0$ does fulfil the required
constraints regarding amplitude and frequency.

\begin{figure}
\centerline{\includegraphics{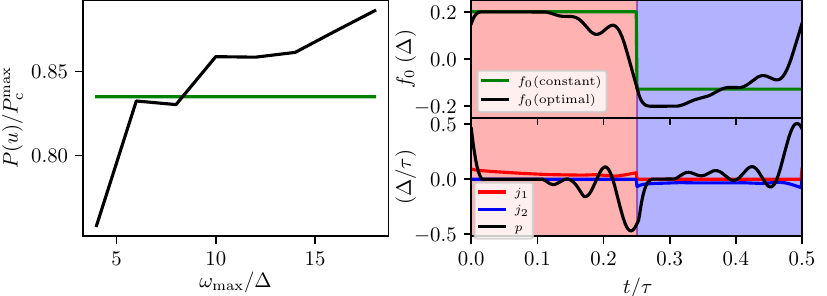}}
\caption{\label{fig:2} {\it Left panel:} Optimized averaged output
  power of the heat engine as a function of the cutoff $\omega_{\rm
    max}$ (black curve), shown as a fraction of the maximum possible power given in Eq.~(\ref{eq:opmaxinfintelyshort}). 
  The
  value obtained when using constant TLS gaps with discontinous fast
  switching is also shown in green.  {\it Top right panel:} Function
  $f_0$ with constant TLS gaps and sudden switching (green), and
  optimized $f_0(\underline{u}^{\rm opt}, t)$. The red and blue shadings mark the
  time regions when the hot and blue baths are connected,
  respectively. {\it Bottom right panel:} Transferred heats and work
  for the heat engine when using $f_0(\underline{u}^{\rm opt}, t)$.  }
\end{figure}

The results shown in Fig~\ref{fig:1} -- in particular, how the
optimised $f_0$ cannot improve the constant gap protocol for very
short $T$ -- point to the relevance of the choice of the cutoff. To
illustrate this fact, we will show the effect of the cutoff in
Fig.~\ref{fig:2}. In this case, the series of runs were done fixing
$T=\frac{1}{2}\tau$, but changing the value of the cutoff frequency,
from $\omega_{\rm max} = 4\Delta$ to $\omega_{\rm max} =
18\Delta$. Increasing the cutoff frequency amounts to enlarging the
search space for the optimisation, and therefore it can be seen on the
left panel how the output power obtained with the optimised $f_0$
increases with $\omega_{\rm max}$. For lower cutoffs, it cannot
improve over the value obtained with the constant gap protocol, but
for larger cutoffs the time-varying optimised $f_0$ permits to obtain
a better number, reflecting the fact that the value obtained with
constant gaps is only a maximum in the limit $T\to 0$. Finally, on the
right hand side of Fig.~\ref{fig:2} we display again function
$f_0(\underline{u}^{\rm opt}, t)$ (top), and the instantaneous heats and work
(bottom), in this case for the calculation with $\omega_{\max} =
18\Delta$. Function $f_0$ changes more rapidly as a function of time
than in the case shown in Fig.~\ref{fig:1}, but it still respects the
constrains imposed on the amplitude and the frequency.

\section{Conclusion}
\label{sec:conclusion}

This work describes and demonstrates a procedure for the optimisation
of the working protocol of QTMs modelled with generic Markovian master
equations. Although there have been numerous works dealing with the
theoretical problem of establishing optimal performance limits for these
systems, there have been few practical, computational methods proposed
in the literature focused on finding realistic optimal protocols for
the action of the external agent or the couplings to the
baths. Specially, if one assumes that ideal unconstrained modes or
operation are not feasible, and bounds on the smoothness or amplitudes
of the control functions are to be considered, based on experimental
or physical considerations. This work intends to fill in that gap.

As a demonstration, we have computed optimised protocols for two-level
systems coupled to thermal reservoirs, that may be for example used to
model simple quantum dots coupled to metallic leads at varying
temperatures. These can operate as minimal quantum heat engines,
producing output power through the intermittent switch from the cold
to the hot bath. The action of the external agent is given by the gap
of the TLS, and it is the shape of this function the one that, in the
example shown, is controlled. It is shown how this function can be
parameterised respecting amplitude and frequency constraints, and how
the value of those parameters can be efficiently optimised using a
gradient-based algorithm. The method can also be used to optimise the
functions controlling the couplings to the baths, internal bath
parameters such as their temperatures if one considers them to be
time-dependent, the duration of the cycle or of each bath coupling,
etc.

There has been a recent push in the research of QTMs, motivated by the
technological and experimental trend towards the miniaturisation of
devices, and by the theoretical questions around the interplay of
thermodynamics for mesoscopic systems and quantum mechanics. I expect
that the method proposed here, and the code that implements it
(published as open source), will be useful to analyse the performance
of these systems.

\section*{Acknowledgements}


\paragraph{Funding information}
AC acknowledges support from Grant PID2021-123251NB-I00 funded by MCIN/AEI/10.13039/501100011033.

\bibliography{qtm.bib}
\bibliographystyle{quantum}

\onecolumn
\appendix

\section{Parameterisation of the control function}
\label{appendix:parametrization}

Function $f_0(\underline{u}, t)$ should fulfill two conditions that would normally
be present in any experimental implementation of a QTM. First, it must be 
bound: $\vert f_0(\underline{u}, t)\vert \le \delta$. Second,
it should be smooth, which we will enforce by also requiring a
high-frequency cutoff. This suggests, for example, the following parametrization:
\begin{equation}
f_0(\underline{u}, t) = \Phi(f_{\rm Fourier}(\underline{u}, t))\,.
\end{equation}
Here, the parameters $\underline{u}$ are the coefficients of a Fourier series:
\begin{equation}
f_{\rm Fourier}(\underline{u}, t) = u_0 + \sum_{n = 1}^{M} \left(
u_{2n} \cos (\omega_n t) + u_{2n-1}\sin(\omega_n t)
\right),\quad (\omega_n = \frac{2\pi}{T}n)\,.
\end{equation}
The function $\Phi$ is chosen with the following properties: (i) for
small $x$, $\Phi(x)\approx x$. Therefore, if $f_{\rm Fourier}(\underline{u}, t)$
is small, $f_0(\underline{u}, t)$ is simply equal to a Fourier series with a
frequency cutoff, $f_{\rm Fourier}(\underline{u}, t)$. (ii) as $\vert x\vert$
grows and approaches $\delta$, the growth of $\vert\Phi(x)\vert$ is
reduced, so that $\vert\Phi(x)\vert \le \delta$ for any $x$, until it
becomes a constant equal to $\delta$ for $x > \delta$.

Many possible functions can be imagined with those properties. For this work, we have chosen:
\begin{equation}
\Phi(x) = \left\{
\begin{array}{l}
x\quad \textrm{ if }0\le x \le \frac{3}{4}\delta
\\
\delta\quad \textrm{ if }x \ge \delta + \frac{1}{4}\delta
\\
\textrm{an Akima cubic spline interpolation if } x \in (\frac{3}{4}\delta, \delta + \frac{1}{4}\delta)
\end{array}
\right.
\end{equation}
For $x<0$, the function should be antisymmetric: $\Phi(x) = - \Phi(-x)$.

This parameterisation strictly enforces the amplitude bound
$\vert\Phi(x)\vert \le \delta$, but it does not enforce the frecuency
cutoff; it only {\it favours} the frequencies lower then $\omega_{\rm
  max}$, as long as the value of the coefficients $u$ are not
high. Because of this, we use a penalty for high
  frequencies in the definition of the merit fucntions.

Many other possible parameterisations of $f_0$ can be imagined; a
particularly simple one would be to use directly $f_0(\underline{u}, t) = f_{\rm
  Fourier}(\underline{u}, t)$, and perform a bound optimisation
setting $\sum\vert u_k\vert \le \delta$. This ensures both the
amplitude and frequency bounds, and has the advantage of its
simplicity -- although that simplicity does not imply a computational
saving. I have also implemented this choice, but found that it
frequently leads to worse local maxima (probably because the optimal
functions found with this parameterisation do not have instantaneous
amplitudes close to the bound $\delta$).

\section{Fluctuations}
\label{appendix:fluctuations}

A good heat engine should not only yield high power, but
also operate with high efficiency and with low power fluctuations. The method described
in Section~\ref{sec:floquetengineering} permits to define merit functions that include the
first two quantities (power and efficiency), but not the third one (fluctuations).
The reason is that the previous method description has only considered observables in the form:
\begin{equation}
\label{eq:observable}
A(\underline{u}) = \frac{1}{T}\int_0^T\!\!{\rm d}t\; {\rm Tr}\left[
\mathcal{M}(\underline{u}, t)(\rho(\underline{u}, t)) N(\underline{u}, t)\,.
\right]\,,
\end{equation}
where $\mathcal{M}$ is a liner superoperator, and $N$ an operator. This includes
the dissipated heats and the power. Therefore, one can define any merit function with the form
$G(\underline{u}) = g(P(\underline{u}), J_1(\underline{u}),\dots, J_B(\underline{u}))$,
where $P$ is the output power and $J_b$ are the dissipated heats to bath $b$ (out of $B$ baths).

However, the form (\ref{eq:observable}) does not allow for the calculation
the power fluctuations. Following Refs.~\cite{Miller2019, Erdman2023}, this
observable can be computed as:
\begin{equation}
\label{eq:deltapdef}
\Delta P(\underline{u}) = \frac{1}{T}\int_0^T\!\!{\rm d}t\; {\rm Tr}\left[
s(\underline{u}, t) \frac{\partial \hat{H}}{\partial t}(\underline{u}, t)
\right]
\end{equation}
Here, $s(\underline{u}, t)$ is a trace-less Hermitian operator that can be
computed from the differential equation:
\begin{align}
\nonumber
\frac{\partial s}{\partial t}(\underline{u}, t) = &
\mathcal{L}(\underline{f}(\underline{u}, t))s(\underline{u}, t)
\\\label{eq:dsdt}
&+ \lbrace \rho(\underline{u}, t), \frac{\partial H}{\partial t}(\underline{u}, t)\rbrace
- 2 {\rm Tr}\left[ \rho(\underline{u}, t)\frac{\partial H}{\partial t}(\underline{u}, t)\right]
\rho(\underline{u}, t)\,.
\end{align}
This equation admits a periodic solution to which $s(\underline{u}, t)$ converges asymptotically~\cite{Erdman2023} -- 
the solution to be used in Eq.~(\ref{eq:deltapdef}). Clearly, $\Delta P$ cannot be brought into
the form given by Eq.~(\ref{eq:observable}).

In order to take into account power fluctuations,
we consider now the possibility of defining a merit function including $\Delta P$:
\begin{equation}
G(\underline{u}) = g(P(\underline{u}), J_1(\underline{u}),\dots, J_B(\underline{u}), \Delta P(\underline{u}))
\end{equation}
Therefore, we need to extend the method in order to be able to compute $\Delta P(\underline{u})$ and its gradient,
\begin{equation}
\label{eq:gradP}
\frac{\partial \Delta P}{\partial u_r}(\underline{u}) = \frac{1}{T}\int_0^T\!\!{\rm d}t\; {\rm Tr}\left[
\frac{\partial s}{\partial u_r}(\underline{u}, t) \frac{\partial \hat{H}}{\partial t}(\underline{u}, t)
\right]\,.
\end{equation}
Fortunately, the formalism can be easily extended for that purpose,
although at the cost of adding some extra computational effort, as it will be shown now.

The necessary ingredients are $s$, and its gradient with respect to $\underline{u}$.
Note that the equation verified by $s$, (\ref{eq:dsdt}), is the same GKSL equation verified by $\rho$,
plus an inhomogeneous term, that depends
non-linearly on $\rho$. Moving it to the spectral representation we get:
\begin{equation}
\label{eq:luh}
\overline{\mathcal{L}}(\underline{u})\tilde{s}(\underline{u}) = h(\underline{u}, \tilde{\rho}(\underline{u}))\,,
\end{equation}
where $h$ is a quadratic function in the $\tilde{\rho}$ components,
that results of the transformation of the inhomogeneous term in 
Eq.~(\ref{eq:dsdt}) to the Fourier series space.
By solving this linear equation for $\tilde{s}(\underline{u})$, one may compute $\Delta P(\underline{u})$,
using Eq.~(\ref{eq:deltapdef}).

Regarding the gradient, one may take derivatives in Eq.~(\ref{eq:luh}),
which leads to the linear equation:
\begin{equation}
\label{eq:dlduds}
\overline{\mathcal{L}}(\underline{u})\frac{\partial \tilde{s}}{\partial u_r}(\underline{u}) =
- \frac{\partial \overline{\mathcal{L}}}{\partial u_r}(\underline{u})\tilde{s}(\underline{u})
+ \frac{\partial}{\partial u_r} h(\underline{u}, \tilde{\rho}(\underline{u}))\,.
\end{equation}
This can be compared to Eq.~(\ref{eq:mainequation}): it is also a linear equation for the 
(super)operator $\overline{\mathcal{L}}(\underline{u})$, although there is an extra term
on the right hand side, that depends quadratically on $\tilde{\rho}$.

Summarising, by solving Eqs.(\ref{eq:luh}) and then (\ref{eq:dlduds}), one may obtain $\tilde{s}$ and
its gradient, that in turn permit to compute $\Delta P$ and its gradient. In this way, the 
fluctuations may be included in the definition of the merit function. Notice that these equations
are similar to the ones used to obtain $\tilde{\rho}$ and its gradient; the extra computational
cost is therefore of similar complexity and scaling.




\end{document}